\documentclass[aps,prl,twocolumn,superscriptaddress,showpacs]{revtex4}

\usepackage{graphicx}

\begin{document}

\title{Fractional $\hbar$-scaling for quantum kicked rotors without cantori}

\author{J. Wang}
\affiliation{Temasek Laboratories,National University of
Singapore,117542 Singapore.} \affiliation{Beijing-Hong
Kong-Singapore Joint Center for Nonlinear and Complex Systems
(Singapore), National University of Singapore,117542 Singapore.}
\author{T.S Monteiro }
\email[]{t.monteiro@ucl.ac.uk}
\affiliation{Department of Physics
and Astronomy, University College London, Gower Street, London,
United Kingdom, WC1E 6BT}
\author{S.~Fishman}
\affiliation{Physics Department, Technion, Haifa, IL-32000, Israel}
\author{ J.P. Keating and R. Schubert}
\affiliation{ School of Mathematics, University of Bristol, Bristol
BS8 1TW, U.K.}
\date{\today}

\begin{abstract}

Previous studies of quantum delta-kicked rotors have found 
 momentum probability distributions  with a typical  width (localization length $L$) 
characterized by fractional $\hbar$-scaling, ie $L \sim \hbar^{2/3}$ 
in regimes and phase-space regions close to
`golden-ratio' cantori. In contrast, in typical chaotic regimes, the
scaling is integer, $L \sim \hbar^{-1}$. Here we consider a 
generic variant of the kicked rotor, the random-pair-kicked particle
(RP-KP), obtained by randomizing the phases every second kick; it
has no KAM mixed phase-space structures, like golden-ratio cantori, at
all. Our unexpected finding is that, over comparable phase-space regions,
 it also has fractional scaling,
but $L \sim \hbar^{-2/3}$. A semiclassical analysis indicates that
the $\hbar^{2/3}$ scaling here is of  quantum origin and is not a
signature of classical cantori.

\end{abstract}


\maketitle

A notable recent achievement of atom optics has been the
realization of a well-known paradigm of quantum chaos, the quantum
`$\delta$-kicked particle' (DKP) \cite{Ott,Reichl}. Laser-cooled
atoms in a pulsed periodic optical potential can be effectively
modeled theoretically by the Hamiltonian $H=p^2/2 - k\cos x \sum_n
\delta (t-nT)$ provided the pulses are sufficiently short. This
system has been extensively investigated in numerous theoretical
(see eg \cite{Casati,Fishman}) and experimental \cite{Moore}
studies by several groups worldwide. In the  chaotic regime ($kT
\gg 1$) the momentum distributions $N(p,t)$ evolve into a final,
time-independent distribution of exponential form: $ N(p, t\to
\infty) \sim \exp (-|p|/L)$. $L$, the momentum localization
length, has well-known integer scaling properties $L \sim K^2
\hbar^{-1}$. This effect, termed `dynamical localization (DL)',
has been well-studied both experimentally and theoretically.

But an important exception, with $L \sim \hbar^{2/3}$, was found in the
 seminal theoretical study of  quantum behavior in the vicinity of
so-called `golden-cantori' in \cite{Geisel}. At a critical value of $kT \approx 0.97$, the last
classical barrier (KAM torus) which impedes chaotic diffusion is
broken. What remains are fractal partial barriers, termed cantori,
situated at momenta $p \approx 2\pi R$ and $p \approx 2\pi (R-1)$,
(or integer multiples thereof) where $R$ is the golden ratio. 
 A subsequent study \cite{Maitra} suggested that
a {\em positive} exponent $L \sim \hbar^{+\sigma}$ was associated with
tunnelling transport (favoured by increasing $\hbar$) while a negative
exponent $L \sim \hbar^{-\sigma}$ was associated with  dynamical
localization where transport increases as $\hbar \to 0$.  In \cite{Maitra}
it was found that the sign of the scaling exponent can change
 from negative to positive as the dominant
transport mechanism changes from tunnelling to dynamical localization.
A  study of the {\em classical} phase-space scaling near
these golden tori \cite{scaling} identified two characteristic scaling exponents,
$\sigma \approx 0.65$ and $\sigma \approx 0.75$.
Since the fractional $L  \sim \hbar^\sigma \sim \hbar^{2/3}$ {\em
quantum} scaling was found in regions close to $p \sim  2\pi R$,
this  behavior was attributed in \cite{Geisel} to the smaller of
the classical exponents.  

A recent study \cite{QKR2} of a closely related system, the double-kicked
particle (2-DKP), found fractional scaling of momentum distributions with
$L \sim \hbar^{-0.75}$,ie  characterized by one of the classical
golden ratio exponents - and in phase-space regions corresponding quite closely to those of
\cite{Geisel}-  it was argued that this too was evidence of
the quantum signature of the golden cantori. As the 2-DKP has
already been experimentally realized with cold atoms \cite{Jones}
and its fractional scaling  occurs over a much broader range of $p$ than
for the usual DKP, this feature of the atomic momentum distributions is very amenable to experimental
study.

However, here, we introduce a system which is random, but closely related 
to both the DKP and 2-DKP: the randomized-pair
DKP or RP-DKP. It is obtained as a limit as the phases between
consecutive kick-pairs become completely random. This system has
no mixed phase-space regions or KAM cantori at all. This study, over
a range of smaller $\hbar$ values than \cite{QKR2}
 found the random system has exactly the same scaling properties as the
mixed-phase-space 2-DKP. Both have $L \sim \hbar^{-2/3}$ for small
$\hbar$; in \cite{QKR2} a slightly larger exponent was found because
some non-asymptotic values of $\hbar$ were included. This unexpected
result is the key finding of our paper. A semiclassical analysis
shows that the scaling with $\hbar^{-2/3}$ arises naturally from the
quantum unitary matrix and is not a signature of golden cantori. We discuss
possible implications for the ordinary DKP.

The Hamiltonian of the 2-DKP is given by \cite{QKR2}
%H= \frac{P'^2}{2 }- k\cos x \sum_N \delta (t'-NT) + \delta (t'-NT+\epsilon) \\
$H= \frac{P'^2}{2 }- k\cos x \sum_N [\delta (t'-NT) + \delta
(t'-NT+\epsilon)] $
where $\epsilon \ll T$ is a small time interval. In effect, the particles
are exposed to a sequence of {\em pairs} of closely spaced $\delta$-kicks.
One could now
adopt the usual procedure and choose
$T$ to define the time unit and thence derive a classical map with
a stochasticity parameter $K=kT$ and momentum units $p = P'T$.
However, we show below that, for the 2-DKP, it is the small
timescale $\epsilon$, rather than the period $T$, which provides
the natural unit of time. Hence, rescaling time as $t'/\epsilon \to t$, and  $
P'\epsilon \to p$, we obtain our 2-DKP map:
\begin{eqnarray}
p_{N+1}=p_N + K_{\epsilon}\sin x_N; &\ & p_{N+2}=p_{N+1} + K_{\epsilon}\sin x_{N+1}
\nonumber \\
x_{N+1}=x_N + p_{N+1}; &\ & x_{N+2}=x_{N+1} + p_{N+2} \ \tau_\epsilon
\nonumber \\
\label{eq3}
\end{eqnarray}
The classical stochasticity parameter $K_{\epsilon}=k \epsilon$.
This map depends also on a further parameter $\tau_{\epsilon}=
(T-\epsilon)/\epsilon$. 

 Clearly, we see that setting $\tau_\epsilon=1$ in Eq.(\ref{eq3}) recovers the
Standard Map or DKP. To obtain  the 2-DKP, we take $\tau_\epsilon \sim 10-100$: 
typical experimental values used in \cite{Jones} are
$\tau_\epsilon \approx 10-25$  and $K_{\epsilon}=0.1$ to $0.5$.
The RP-DKP is obtained by taking the limit $\tau_\epsilon \to
\infty$, causing the impulse $\sin x_{N+2}$ to become randomized.
This can be achieved in practice by keeping $k$ and $\epsilon$
constant and taking $T \to \infty$. In the usual $T$-scaled map
this would yield an infinite stochasticity constant $K=kT$ (but
little insight). This  RP-DKP limit has no mixed phase space
behavior at all, but it retains the momentum trapping and its
behavior is determined by the stochasticity parameter
$K_\epsilon$. The map Eq.(\ref{eq3}) has $2\pi$ periodicity in these re-scaled
momentum units for integer $\tau_\epsilon$. 

\begin{figure}
\vspace{-.8cm}
\includegraphics[width=4.in,clip=true]{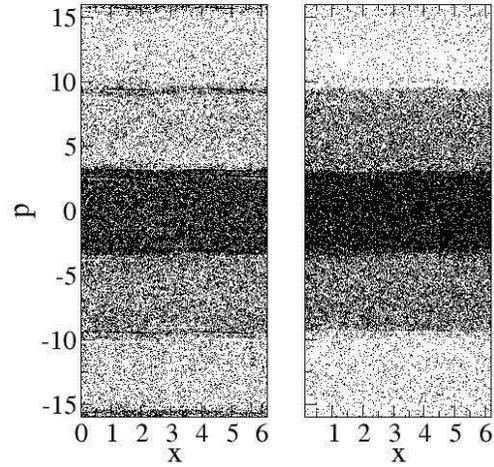}
\vspace{-1.cm} \caption{Comparison of the Poincar\'e SOS for the
2-DKP and Randomised-Pair DKP. Stochasticity parameter
$K_{\epsilon}=0.4$. {\bf (Left)} 2-DKP: $\tau_\epsilon=24$. A
periodic structure of chaotic `cells' separated by thin
mixed-phase regions (coinciding with the trapping regions) is
apparent. {\bf(Right)} RP-DKP: $\tau_\epsilon=10^{6}$ so impulses
between kick pairs are uncorrelated. A similar plot is obtained by
taking $x_{N+2}$ to be a random number. The strength of trapping
is largely unchanged, but all remnants of mixed-phase space
structures are eliminated. Our key finding here is that the
fractional scaling of the {\em quantum} localization lengths
remains unchanged for the RP-DKP.} \label{Fig1}
\end{figure}

Fig.\ref{Fig1} (a) compares Poincar\'e surfaces of section (SOS)
for the 2-DKP and RP-DKP. For the 2-DKP, a
periodic structure of chaotic `cells' separated by thin
mixed-phase regions is apparent. These momentum `trapping' regions
appear at $p \approx (2m+1)\pi$ where $m=0,\pm 1,\pm 2...$. For
odd-integer multiples of $\pi$ there is near-cancellation of
consecutive kicks. Fig.\ref{Fig1} (b) shows the RP-DKP, for which
$\tau_\epsilon = 10^6$. In fact, an indistiguishable SOS can be
obtained by taking actual random phases, i.e. $x_{N+2} =2\pi
\zeta_{N}$ where $0<\zeta_{N}<1$ is a random number chosen every
kick pair. We see that the trapping is there, but with no trace of
islands etc. The RP-DKP never has any islands, regardless of how
small $K_\epsilon$ becomes.

For the quantum equivalent we note that in a basis of plane waves,
the one-period time-evolution operator for the 2-DKP,
$\hat{U}^{\epsilon}$, has matrix elements $
U_{lm}^{\epsilon}=U_l^{free}. \ U_{lm}^{2kick}$ where
\begin{eqnarray}
U_{lm}^{2kick} = \sum_k J_{l-k}\left(K_\hbar\right)
J_{k-m}\left(K_\hbar \right) \exp\left(-i
{k^2\hbar_{\epsilon}}/{2}\right) \label{eq4}
\end{eqnarray}
where $J_n$ is the integer Bessel functions of the first kind,
$\hbar_\epsilon=\hbar\epsilon$ is the effective value of Planck's
constant.
 $K_\hbar=K_\epsilon/\hbar_\epsilon$ and $U_{lm}^{free}(\tau_\epsilon)= i^{l-m} \exp\left(-i {l^2}
\tau_{\epsilon}\hbar_{\epsilon}/{2}\right)$.
 To within an unimportant phase, it can be shown that \cite{QKR2}:
\begin{eqnarray}
U_{lm}^{2kick}\simeq J_{l-m}\left( \frac{2K_\epsilon}{\hbar_\epsilon}\cos \left [l\hbar_{\epsilon}/2\right] \right)
\label{eq5}
\end{eqnarray}
Eq.\ref{eq5} can be derived most straightforwardly by evaluating the sum in Eq.\ref{eq4} using Poisson summation.
The key point is that $U_{lm}^{2kick}$ is common to all the kicked systems: only $U_{lm}^{free}(\tau_\epsilon)$
determines whether we investigate the standard DKP ($\tau_\epsilon=1$), the
mixed phase-space 2-DKP ($1 < \tau_\epsilon \lesssim 100$) or the random
RP-DKP ($\tau_\epsilon \to \infty$).
 Equivalently, we can also obtain the RP-DKP by taking $U_l^{free} =\exp\left(-i 2 \pi \zeta_l\right)$,
 i.e. using random phases for each angular momentum $l$. Taking
$\tau_\epsilon=10^4 - 10^6$ gives the same behavior,
provided  $\tau_{\epsilon}$ is not a rational multiple of $\pi$: transport can be strongly affected at
such resonances.

\begin{figure}
\vspace{-.1cm}
\includegraphics[width=0.37\textwidth,clip=true]{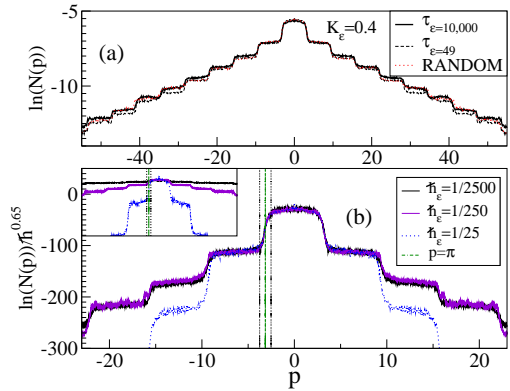}
\caption{{\bf (a)} Shows that the quantum momentum distributions for
the mixed phase-space double kicked atoms (2-DKP) are similar to those of
the RP-DKP, for which the phases between kick pairs are random
numbers. The staircase structure and `heights' of the steps are
comparable. $K_\epsilon=0.4$ and $\hbar_\epsilon=1/500$ in every
case. {\bf (b)} Shows the fractional scaling for the 2-DKP.
$K_\epsilon=0.4$ and $\tau_\epsilon=49$.
 The $\hbar_\epsilon^{2/3}$ rescaling of the
momentum distributions shows that the first step scales
near-perfectly in the range from $\hbar_\epsilon=1/2500$ to
$1/25$; i.e. the `local' localization length in the trapping
region scales over this entire $\hbar_\epsilon$ range. The global
localization length (the envelope of the whole distribution) shows
this fractional $\hbar_\epsilon$ scaling holds only for the
smaller $\hbar_\epsilon =1/2500$ and $\hbar_\epsilon=1/250$ but
not for $\hbar_\epsilon=1/25$. Vertical lines indicate width and
centre ($p=\pi$) of the first trapping region. The inset shows the same curves
without the $\hbar^{2/3}$ scaling.} \label{Fig2}
\end{figure}

Fig.\ref{Fig2}(a) shows the asymptotic quantum momentum distributions,
$N(p, t \to \infty)$, obtained from an initial state $\psi(t=0) \sim
\delta(p-p_0)$ for $p_0=0$, in  regimes analogous to
Fig.\ref{Fig1}. The
dashed line corresponds to a momentum distribution obtained for
the mixed phase-space 2-DKP while the solid line represents
the equivalent result obtained using random phases in $U_l^{free}$.
One sees that both the mixed phase-space 2-DKP and the random
map (RP-DKP) distributions are very similar, with a distinctive `staircase'
structure.

\begin{figure}
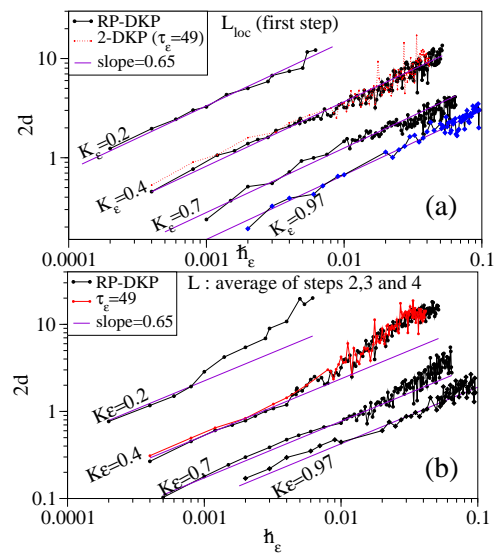

%\begin{figure}[htb]
%\hspace*{1 mm}
\includegraphics*[width=2.5in]{Fig3a.eps}
\vspace*{10 mm}
\includegraphics[width=2.5in]{Fig3b.eps}
\vspace*{-10 mm} \caption{{\bf(a)} Fractional scaling of the
$L_{loc}$ (the localization length at the first step: note that
$L_{loc} \sim (2d)^{-1}$). It shows the fractional $2d \sim
\hbar_\epsilon^{2/3}$ scaling holds over the full
$\hbar_\epsilon$ range. Straight lines indicate slope $0.65$.
The $\approx 2/3$ scaling holds even for $K_\epsilon=0.97\approx K_c$, the
parameter for criticality in the Standard Map: the only difference is that here
 $\tau_\epsilon \to \infty$ (RP-DKP) instead of $\tau_\epsilon=1$ (Standard Map).
{\bf(b)} As for (a) but shows the average of steps 2 to 4, showing
that the $L \sim \hbar_\epsilon^{-2/3}$ global scaling persists only
for the smaller $\hbar_\epsilon $ values.}
\label{Fig3}
\end{figure}

Fig.\ref{Fig2}(b) shows the staircases obtained for
$K_\epsilon=0.4$, $\hbar_\epsilon=1/25,1/250$ and
$\hbar_\epsilon=1/2500$ for the 2-DKP. We have rescaled the
momentum distributions $N(p)$ using the `local' scaling exponent
estimated in Fig.\ref{Fig3} for $K_\epsilon=0.4$ which yielded
$\sigma\approx 0.65$ . We plot ln$(N(p))\hbar^{-0.65}$, shifted
 by an appropriate constant. The figure shows that the shape and magnitude of the
distribution around the first trapping region re-scales perfectly
with $\hbar_\epsilon$. $L_{loc}$ is the slope on the first step.
The height of the steps ($2d$ in the notation of \cite{QKR2})
scales fractionally with $\hbar_\epsilon$, while the width $w$
(momentum width between the dashed vertical lines) is independent
of $\hbar_\epsilon$: in scaled momentum units $w \sim 2\pi/6$.
Thus we take $L_{loc} \approx w/2d $, i.e. for the scaling,
$L_{loc} \sim (2d)^{-1}$.

The longer-ranged localization length $L$, characterizing the
envelope of the full staircase, scales well in the smaller
$\hbar_\epsilon$  range $\hbar_\epsilon \lesssim
1/250$ but we note that the $\hbar_\epsilon=1/25$ staircase does
not scale at long-range.
%Such fractional
%$\hbar_\epsilon$ for the 2-DKP might not be too
%surprising: a detailed study of the thin mixed-phase space regions
%\cite{Mischa} showed that locally, the resonance structure is similar to
% the Standard Map's. However, this scaling also persists for
%the RP-DKP, where it is rather harder to understand.

In Fig.\ref{Fig3} we investigate the
$\hbar_\epsilon$ scaling exponent itself. For staircases where
 $\hbar_\epsilon$ is sufficiently small,
ln$(N(p)) \approx const$ outside the `steps' in the trapping
regions. Then, only the step heights depend on
$\hbar_\epsilon$. We can restrict ourselves to a study of the
parameter $2d$ since it has the same
$\hbar_\epsilon$ scaling as $L_{loc}$. The midpoint of the
$n$-th step is at $p_{1/2}^{(n)}=2(n-1)\pi/\epsilon$ for $n\neq
1$; for $n=1$, we take $p_{1/2}^{(1)}=\pi/(2\epsilon)$. For the
$n$-th step we measure
\begin{eqnarray}
2d(n)=\langle ln(N(p_{1/2}^{(n)}))\rangle - \langle
ln(N(p_{1/2}^{(n+1)}))\rangle
\end{eqnarray}
where the average is taken over a small momentum interval around
$p_{1/2}^{(n)}$. This procedure is easily automated, allowing a
very fine-grid of $\hbar_\epsilon$ values, so
we can examine fluctuations in behavior.

\begin{figure}[htb]
\includegraphics*[height=2.0in]{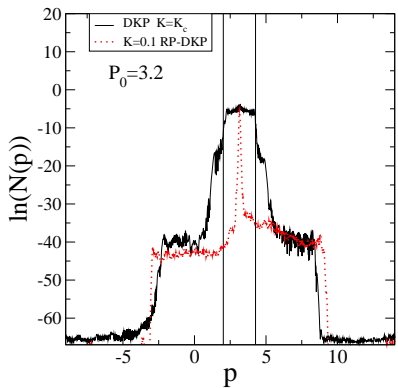}
\caption{Figure shows that the RP-DKP  trapping regions
coexist with the phase space regions which contain the
two `golden ratio' tori of the DKP (Standard Map); nevertheless,
destroying the cantori in the RP-DKP with random phases does not
eliminate  fractional scaling. (Bold solid
lines) DKP at critical $K_c \approx 0.97$; the plot is similar to
Fig.2b of Geisel et al 1986 \cite{Geisel} which yielded $L_{loc}
\sim \hbar^{+2/3}$. Vertical lines indicate `golden' tori. Graph
was obtained by setting
$K_\epsilon=K_c=0.97$, $\tau_\epsilon=1$, $\hbar_\epsilon=0.01$.
 (Dotted line) the RP-DKP distribution
is shown for a kick strength $K_\epsilon=0.1$,
 sufficiently small that the amplitude `drops' are comparable
in magnitude to the DKP (but without the golden cantori). }
\label{Fig4}
\end{figure}

Fig.\ref{Fig3}(a) compares the
$\hbar_\epsilon$ scaling of $2d(1)$ for the 2-DKP and the
RP-DKP, while Fig.\ref{Fig3}(b) shows the average of $2d$ for
steps $n$=2 to 4. We take the behavior of the first step to
indicate the local localization length and of steps 2 to 4 to
indicate the properties of the long-ranged scaling. All the plots,
whether 2-DKP or RP-DKP, show a slope $\sigma \approx 0.65$ for
small $\hbar_\epsilon$. For the upper ranges of $\hbar_\epsilon$,
the slope (for $L$ only) increases significantly. These results are quite consistent
with the slightly higher estimate $\sigma \approx 0.75$ found in
\cite{QKR2}: there, an average of the first few steps was
obtained, including a few $\hbar_\epsilon$ values in the range where
the deviation from $\sigma \approx 2/3$ begins.

We note that the trapping regions and the Standard Map golden
cantori regions occupy similar regions of phase-space: the golden
tori occur at winding number $p_0/2\pi \approx 0.618$ and
$p_0/2\pi \approx (1-0.618)$. The trapping occurs for $p_0/2\pi
\approx 0.5$ (e.g. momentum $p \approx \pi$), exactly midway
between these. In Fig.(\ref{Fig4}) we show the well-known plot of
Geisel et al (Fig.2(b) in \cite{Geisel}) reproduced in quantum
chaos textbooks \cite{texts}. In that study, an initial state
 $\psi(t=0)\sim \delta(p-p_0)$ was evolved for
critical $K_c\approx 0.9716$ to study transport through the golden
cantori. Notably, that study used precisely $p_0= 3.2 \approx \pi$:
 the initial state was centred in the trapping region. We compare
a RP-DKP result also with $p_0=3.2 \approx \pi$ but destroying all
the KAM structures including cantori by taking $\tau_\epsilon \to
\infty$ but otherwise running exactly the same numerical code (We
take the same $\hbar=1/100$ but $K=0.1$ for the RP-DKP
to keep diffusion rates comparable to the DKP).
Since $\epsilon$ is our unit of time in both systems,  we can
drop the $\epsilon$ subscripts on $\hbar$ and $K$.

We now analyze the cause of the $L \sim \hbar^{-2/3}$ scaling of the RP-DKP.
 We consider the
trapping regions $\delta p=(p-p_0)$ centered at $p_0= l_0
\hbar=(2n+1) \pi$. At the centers of these regions the argument of
the Bessel function in Eq.\ref{eq5} vanishes.  Near these points it
is of appreciable size up until when its argument
$x=\frac{2K}{\hbar}\cos \left[l\hbar/2\right]$ is of the order of
the index; i.e.~when $x\lesssim|l-m|$. If  also $|l-m| \gg 1$, the
asymptotic approximation
\begin{equation}
U^{2kick}_{lm} \sim J_{l-m} (x) \sim \frac{2^{1/3}}{|l-m|^{1/3}}{Ai(-2^{1/3}z}),
\end{equation}
 with $x=|l-m|+\frac{z}{|l-m|^{1/3}}$, holds. $\delta p$ is small,
as we are near the center of a bottleneck, so
this implies we are in the semiclassical, $\hbar \to 0$, regime.
Heuristically, the transition probability corresponding to a small
change in momentum $\Delta p$ is $|U_{lm}^{2kick}|^2$, which is
proportional to $|l-m|^{-2/3}=\frac{\hbar^{2/3}}{\Delta p^{2/3}}$
(where $\Delta p=|l-m|\hbar$). The effect of $U_{lm}^{free}$ is to
randomize the phases. The trapping regions act as
bottlenecks for transport. On scales larger than $\Delta p$,
considered fixed, diffusion takes place: $p^2 \sim Dt$, where the
local diffusion coefficient $D$ is proportional to
$|U_{lm}^{2kick}|^2$ and therefore to $\hbar^{2/3}$. Assuming that
the evolution operator is a band-matrix with constant width (as is
the situation for the ordinary Kicked-Rotor), the local localization
length satisfies \cite{Shep, Fishman} $ L_{local}\sim\frac{D}{\hbar}$,
implying  $L_{local} \sim \hbar^{-1/3}$. In our case, the width of the
contributing region, as implied by the scaling of the argument of
the Airy function in Eq. (6), is of the order
$|l-m|^{1/3}=\frac{\Delta p^{1/3}}{\hbar^{1/3}}$. This should
multiply the expressions for $D$ and $L_{local}$. The resulting
localization length therefore satisfies $L \sim \hbar^{-2/3}$.  The
condition for the validity of the asymptotic properties of the
Bessel function and for the momentum to be in a bottleneck region is
$1 \gg \cos \left[l\hbar/2\right] \gg \frac{\hbar}{2K}$.

We recall that this structure of $U^{2kick}_{lm}$ is common to both the quantum DKP as well
as RP-DKP; and Fig.\ref{Fig4} shows that the fractional scaling, is observed
over similar phase-space regions. One may safely conclude from the analysis here that the
behaviour of the generic RP-DKP follows only from the semiclassical dynamics in the
bottleneck regions: it is the hallmark of the scaling of an Airy
function rather than of the fractal scaling near golden cantori.
However, for the DKP, the additional presence of large stable islands bordering
the regions where $\hbar^{2/3}$ is observed
 (which occur when $\tau_\epsilon \approx 1$) make the quantum-classical 
interplay less transparent. In this case too, Airy functions and $\hbar^{ 2/3}$ factors
occur naturally in the semiclassical quantization of torus states \cite{Berry}.
A much more detailed study than the one undertaken here
is required to conclusively establish whether the semiclassical
dynamics is also the dominant mechanism in the quantum DKP,
but we suggest that cantori do not represent the only possible source
for  $\hbar^{\pm 2/3}$ scaling behavior.

The authors thank C.Creffield and A. M.
Garc\'{\i}a-Garc\'{\i}a for helpful discussions. J.W. acknowledges
support from Defence Science and Technology Agency (DSTA) of
Singapore under agreement of POD0613356.

\end{document}